\documentclass[12pt]{article}
\usepackage{latexsym, amsmath, amssymb}
\newtheorem{thm}{Theorem}
\newtheorem{cor}{Corollary}
\newtheorem{lem}[thm]{Lemma}
\newtheorem{prop}[thm]{Proposition}
\newtheorem{defn}{Definition}
\newtheorem{rem}{Remark}

\newcommand{\set}[1]{\left\{#1\right\}}
\newcommand{\Real}{\mathbb{R}}
\newcommand{\Complex}{\mathbb{C}}
\newcommand{\To}{\longrightarrow}
\newcommand{\image}[1]{\textrm{Im}(#1)}
\newcommand{\kernel}[1]{\textrm{ker}(#1)}

\newenvironment{prf}{\textbf{Proof.}}{$\blacksquare$}
\newcommand{\pars}[1]{\left(#1\right)}
\newcommand{\jami}[2]{\sum\limits_{#1}^{#2}}
\newcommand{\all}[1]{\forall\,#1}
\newcommand{\vbar}{\Bigr|}
\newcommand{\imply}{\Rightarrow}
\begin{document}

\title{The Characteristic Class of a Lie Algebra Ideal, Contact Structures and the Poisson Algebra of Basic Functions}%
\author{Zakaria Giunashvili\\
\small{Department of Theoretical Physics,}\\
\small{Institute of Mathematics, Georgian Academy of Sciences}\\
\small{Tbilisi, Georgia}\\
\small{e-mail: zaqro2002@hotmail.com}}

\date{\today}
\maketitle
\begin{abstract}
Some cohomology classes associated with an ideal in a Lie algebra, the Poisson structure on
the algebra of basic functions for a contact structure, its Poisson cohomologies and geometric
(pre)quantization are considered and investigated from the algebraic point of view.
\end{abstract}
\newpage
\tableofcontents
\newpage
\section[Introduction]{\large{Introduction}}
A simple situation when we have a Lie algebra $L$ wich is a module over a commutative and associative algebra $A$,
and an ideal $V$ in it, gives rise to many interesting structures. In this paper we consider some
generalization of the well-known characteristic
classes of a fiber bundle from this point of view. This classes, in this case, are the elements
of the cohomology space of $L$ with values in the homology space of $V$. Also we define a Poisson
structure on the basic algebra (i.e. the algebra of such elements $a\in A$ that $V(a)=0$)
and then consider the case of the contact manifold from this point of view. Investigate the Poisson
cohomologies of this Poisson structure and its prequantization.
\section[Lie Algebra Cohomologies with Values in a Module]{\large{Lie Algebra Cohomologies with Values in a Module}}\label{section1}
We start from review of some definitions and facts from the cohomology theory of Lie algebras.
Let $L$ be a Lie algebra, $A$ be an associative and commutative algebra over the field of complex (or real)
numbers and $S$ be a module over the algebra $A$.
\begin{defn}\label{Lie_module_definition}
The triple $(L,S,A)$ is said to be a Lie module over the Lie algebra $L$ if the following conditions
are satisfied:
\begin{itemize}
\item[\textbf{(l1)}]
$L$ is a module over the algebra $A$ and there is a Lie algebra homomorphism $\phi:L\To Der(A)$,
which is also a homomorphism of $A$-modules, such that for any $a\in A$ and $X,Y\in L$, we have:
$[X,aY]=X(a)Y+a[X,Y]$. Here $Der(A)$ denotes the space of derivations of the algebra $A$ and
$X(a)$ denotes $\phi(X)(a)$.
\item[\textbf{(l2)}]
There is a Lie algebra homomorphism $\psi:L\To End_\Complex(S)$, which is also a homomorphism of
$A$-modules, such that for any $X\in L,\;s\in S$ and $a\in A$, we have: $X(as)=X(a)s+aX(s)$.
Here $End_\Complex(S)$ denotes the space of $\Complex$-linear mappings from $S$ to itself and $X(s)$
denotes $\psi(X)(s)$.
\end{itemize}
\end{defn}
Sometimes, for brevity, the term ``Lie module over $L$'' will be used for the $A$-module $S$.

For any integer number $m\geq1$, let us denote the space of $A$-multilinear mappings from
$\underbrace{L\times\cdots\times L}_{m-times}$ to the Lie module $S$, by $\Omega_A^m(L,S)$. We also set that
$\Omega_A^0(L,S)=S$ and $\Omega_A(L,S)=\bigoplus\limits_{m=0}^{\infty}\Omega_A^m(L,S)$.

For any $\omega\in\Omega_A^m(L,S)$ and $\set{X_1,\cdots,X_{m+1}}\subset L$, define $d\omega$
by the well-known Koszul formula:
$$
\begin{array}{l}
(d\omega)(X_1,\cdots,X_{m+1})=\jami{i}{}(-1)^{i-1}X_i\omega(X_1,\cdots,\widehat{X_i},\cdots,X_{m+1})+\\
+\jami{i<j}{}(-1)^{i+j}\omega([X_i,X_j],\cdots,\widehat{X_i},\cdots,\widehat{X_j},\cdots)
\end{array}
$$
\begin{lem}\label{form_differential_is_form_lemma}
If the $A$-module $L$ is projective then the following three conditions are equivalent:
\begin{enumerate}
\item
For any $a\in A$ and $s\in\Omega^0_A(L,S)=S$: $d(as)=d(a)s+ad(s)$.
\item
The $A$-module $S$ is a Lie module over $L$.
\item
For any integer $m\geq1$ and $\omega\in\Omega^m_A(L,S)$, the mapping
$$d\omega:L^{m+1}\To S$$
is $A$-multilinear (i.e., $(d\omega)(aX_1,\cdots,X_{m+1})=a(d\omega)(X_1,\cdots,X_{m+1})$).
\end{enumerate}
\end{lem}
\begin{prf}
It is clear that the conditions 1 and 2 are equivalent, because for any $X\in L$, we have that:
$$
\pars{d(as)}(X)=X(as)=X(a)s+aX(s)=\pars{d(a)s+ad(s)}(X).
$$
Also, it can be verified by direct calculation that from the condition 2 (or, which is the same,
from 1) follows the condition 3. Now, suppose that the condition 3 is true. In this case For any
$X,Y\in L,\;a\in A$ and $\omega\in\Omega_A^1(L,S)$, we have:
$$
\begin{array}{l}
(d\omega)(X,aY)=a(d\omega)(X,Y)\quad\Rightarrow\\
\\
X\omega(aY)-aY\omega(X)-\omega([X,aY])=aX\omega(Y)-aY\omega(X)-a\omega([X,Y])\;\Rightarrow\\
\\
X(a\omega(Y))-aY\omega(X)-X(a)\omega(Y)-a\omega([X,Y])=\\
\\
=aX\omega(Y)-aY\omega(X)-a\omega([X,Y])\qquad\Rightarrow\\
\\
X(a\omega(Y))=X(a)\omega(Y)+aX(\omega(Y))
\end{array}
$$
Because it is assumed that the $A$-module $L$ is projective, for any $s\in S$ can be found such
$\omega\in\Omega^1_A(L,S)$ and $Y\in L$, that $\omega(Y)=s$. Therefore, we obtain that
$X(as)=X(a)s+aX(s)$, i.e., $S$ is a Lie module over the Lie algebra $L$.
\end{prf}

Hence, we have that if the $A$-module $S$ is a Lie module over the Lie algebra $L$, the operator
$d$, carries the space $\Omega^m_A(L,S)$ into $\Omega^{m+1}_A(L,S)$, which implies that the pair
$(\Omega_A(L,S),d)$ is a differential complex.
\section[Lie Algebra Homologies and Supercommutator]{\large{Lie Algebra Homologies and Supercommutator}}\label{section2}
\newcommand{\g}{\ensuremath{\mathfrak{g}} }
Let \g be a Lie algebra and a module over a commutative and associative algebra $B$. If the Lie
algebra bracket in \g is bilinear for the elements of $B$ ($[x,ay]=a[x,y],\;\all{x,y}\in\g$ and
$a\in B$) one has a well-defined boundary operator $\delta:\wedge_B^m(\g)\To\wedge_B^{m-1}(\g)$,
where $\wedge_B$ denotes the exterior product of a $B$-module by itself:
$$
\begin{array}{l}
\delta(x_1\wedge\cdots\wedge x_m)=\jami{i<j}{}(-1)^{i+j}[x_i,x_j]\wedge\cdots\wedge\widehat{x_i}\wedge\cdots\wedge\widehat{x_j}\wedge\cdots\wedge x_m,\;m>1\\
\textrm{ and }\quad\delta(x)=0,\;\all{\set{x,x_1,\cdots,x_m}}\subset\g
\end{array}
$$
The homology given by the boundary operator $\delta$ is known as the homology of the Lie algebra
\g with coefficients from the algebra $B$.

There is a useful relation between the coboundary operator $\delta$ and the Schouten-Hijenhuis
bracket on the exterior algebra $\wedge_B(\g)$: $\all{u}\in\wedge_B^m(\g)$ and
$\all{v}\in\wedge_B(\g)$, we have
\begin{equation}\label{super_as_oper_deviation_formula}
[u,v]=\delta(u)\wedge v+(-1)^mu\wedge\delta(v)-\delta(u\wedge v)
\end{equation}
It can be said that the supercommutator $[\cdot,\cdot]$ measures the deviation of the boundary
operator $\delta$ from being an antidifferential of degree -1 (see \cite{Me}). From the formula
\ref{super_as_oper_deviation_formula}, easily follows that the induced supercommutator on the homology
space $H_{\bullet}(\g,B)$ is trivial: for $u\in\wedge_B^m\g$ and $v\in\wedge_B\g$, such that
$\delta(u)=\delta(v)=0$, we have $[u,v]=-\delta(u\wedge v)$, which implies that the homology class
of the element $[u,v]$ is trivial.

From the formula \ref{super_as_oper_deviation_formula}, also follows that the homology space
$H_\bullet(\g,B)$ does not inherits the exterior algebra structure from $\wedge_B(\g)$: if the
elements $u,v\in\wedge_B(\g)$ are closed, then we have that $\delta(u\wedge v)=-[u,v]$.
\section[The Homology Space of a Lie Algebra Ideal]{\large{The Homology Space of a Lie Algebra Ideal}}\label{section3}
Let $L$ be a Lie algebra and $A$ be a commutative and associative algebra over $\Complex$ (or $\Real$). Suppose
that the pair $(L,A)$ satisfies the condition \emph{l1} from the definition \ref{Lie_module_definition}.

Let $V\subset L$ be an ideal in the Lie algebra $L$ (i.e., $\all{v}\in V$ and $\all{x}\in L$:
$[v,x]\in V$). Define the subalgebra $A_V$, in the algebra $A$, as
$$
A_V=\set{a\in A\,|\,v(a)=0,\;\all{v}\in V}
$$
and the subspace $V'$ in $L$, as
$$
V'=\set{\jami{i=1}{n}a_iv_i\,\vbar\,n\in\mathbb{N}\;a_i\in A_V,\;v_i\in V,\;i=1,\cdots,n}
$$
The latter can be defined as the minimal $A_V$-submodule of $L$, containing $V$.

For any $x\in L,\;v\in V$ and $a\in A_V$, we have
$$
v(x(a))=[v,x](a)+x(v(a))=0\quad\imply\quad x(a)\in A_V
$$
which implies that the subalgebra $A_V$ is invariant under the action of the elements of $L$.
Therefore, we have that $[x,av]=x(a)v+a[x,v]\in V'$, which means that $V'$ is also an ideal in the
Lie algebra $L$. We call the ideal $V'$ the \emph{complement} of the ideal $V$.

From the fact that $V$ is a subspace of $V'$ follows that $A_{V'}\subset A_V$, but on the other
hand, for any $a,b\in A_V$ and $v\in V$, we have that $(bv)(a)=b\cdot v(a)=0$, which means that
any $a\in A_V$ is also an element of $A_{V'}$. Hence, the two algebras $A_V$ and $A_{V'}$, coincide.

The ideal $V$ in the Lie algebra $L$ will be called \emph{complete}, if $V=V'$. Further, by default,
we assume that the ideal $V$ is complete (i.e., $V$ is a module over the the algebra $A_V$).

Consider the homology space of the Lie algebra $V$ with coefficients from $A_V$. One can define
an action of the Lie algebra $L$ on the $A_V$-module $H_\bullet(V,A_V)$: for $X\in L$ and
$v\in V$ such that $\delta(v)=0$, let $X([v])=[[X,v]]$, where $[\cdot]$ denotes the homology class
and $[\cdot,\cdot]$ denotes the supercommutator.
\begin{prop}
The action of the Lie algebra $L$ on the homology space\\
$H_\bullet(V,A_V)$ is correctly defined and gives a Lie module structure on the\\
$A_V$-module $H_\bullet(V,A_V)$.
\end{prop}
\begin{prf}
As it follows from the formula \ref{super_as_oper_deviation_formula}, we have that for any
$X\in L$ and a closed element $u\in\wedge_{A_V}(V)$: $[X,u]=-\delta(X\wedge u)$, which implies that
$[X,u]$ is closed. If the element $u$ is exact and $u=\delta(v)$, then we have
$$
\delta([X,v])=\delta(-X\wedge\delta(v)-\delta(X\wedge v))=-\delta(X\wedge u)=-[X,u]
$$
which implies that the element $[X,u]$ is exact. From these two fact follows that the
closed and exact element in the exterior algebra $\wedge_{A_V}(V)$ are invariant under the supercommutator
with the elements of the Lie algebra $L$. Therefore, the action, $X([u])=[[X,u]]$, of $L$ on the homology
space $H_\bullet(V,A_V)$ is correctly defined.
By definition of the algebra $A_V$ and the properties of the supercommutator, easily follow that
for any $a\in A_V$: $(aX)([u])=aX([u])$ and $X(a[u])=X([au])=[[X,au]]=X(a)[u]+aX([u])$; which
implies that the $A_V$-module $H_\bullet(V,A_V)$ is a Lie module over the Lie algebra $L$.
\end{prf}\bigskip\\
\newcommand{\homid}[3][]{\Omega_{A_V}^{#1}(#2,\;H_{#3}(V,A_V))}
\newcommand{\hmid}[3][]{\Omega_{A_V}^{#1}(#2,H_{#3}(V,A_V))}
It follows from the above proposition, that for each integer $n\geq0$, one can consider the
the following differential complex $\pars{\homid{L}{n},\;d}$, which gives the Lie algebra cohomology
of $L$ with values in the $A_V$-module $H_n(V,A_V)$.

As it was mentioned early, the induced supercommutator on each homology space $H_n(V,A_V),\;n=0,\cdots,\infty$,
is trivial. Therefore, the action of the Lie algebra $V$ on each $H_n(V,A_V)$ is trivial, which
implies that the submodule of the $A_V$-module $\homid{L}{\bullet}$, consisting of such forms
$\omega$, that $i_v(\omega)=0,\;\all{v}\in V$, is invariant under the action of the differential
$d$ ($i_v\omega=0\;\imply\;i_v(d\omega)=0,\;\all{v}\in V$). Hence, one can consider the subcomplex
$\pars{\homid{L}{\bullet}_0,\;d}$ of the differential complex $\pars{\homid{L}{n},\;d}$, where
$$
\homid{L}{\bullet}_0=\set{\omega\in\homid{L}{\bullet}\;|\;i_v(\omega)=0,\;\all{v}\in V}
$$
Actually, the complex $\pars{\homid{L}{\bullet}_0,\;d}$ is canonically isomorphic to the complex $\pars{\homid{L/V}{\bullet},\;d}$.
\section[The Characteristic Class of a Lie Algebra Ideal]{\large{The Characteristic Class of a Lie Algebra Ideal}}
Let us denote by $\homid{L}{\bullet}_1$ the quotient module
$$\homid{L}{\bullet}/\homid{L}{\bullet}_0$$
For any integer $n\geq 0$, we have the following short exact sequence:
$$
\begin{array}{ll}
0\;\To\;\homid{L/V}{n}\;\To\;\homid{L}{n} & \stackrel{\pi}{\To} \\
 \stackrel{\pi}{\To}\;\homid{L}{n}_1\;\To\;0 &
\end{array}
$$
which induces the standard homomorphism of cohomology spaces
$$
\sigma_n:H^\bullet(L,\;H_n(V,A_V))_1\To H^{\bullet\,+1}(L/V,\;H_n(V,A_V))
$$
where $H^\bullet(L,\;H_n(V,A_V))_1$ denotes the cohomology space of the complex $(\homid{L}{n},\;d)$
and $\sigma_n$ is defined as: $\sigma_n([\pi(\omega)])=[d\omega]$, for $\omega\in\homid{L}{n}$,
such that $d(\pi(\omega))=0$ ($\;\imply\;\pi(d\omega)=0\;\imply\;d\omega\in\homid{L/V}{n}$);
here $[d\omega]$ denotes the cohomology class of the form $d\omega$ in $H^{\bullet\,+1}(L/V,\;H_n(V,A_V))$.

\newcommand{\cohom}[2][]{\Omega_{A_V}^{#1}(#2,\,H_V)}

Consider the special case when $n=1$. Let us denote the homology space $H_1(V,A_V)$ by $H_V$. So,
we have an exact sequence
\begin{equation}\label{cohomology_exact_seq2}
0\;\To\;\cohom[\bullet]{L/V}\;\To\;\cohom[\bullet]{L}\;\stackrel{\pi}{\To}\;\cohom[\bullet]{L}_1\;\To\;0
\end{equation}
and the homomorphism: $\sigma_1:H^{\bullet}(L,H_V)_1\To H^{\bullet\,+1}(L/V,H_V)$.

There is a one-to-one correspondence between the set of splittings of the following short exact
sequence
$$
0\;\To\;V\;\To\;L\;\stackrel{p}{\To}\;L/V\;\To\;0
$$
and the set of such homomorphisms $\alpha:L\To V$, that $\alpha(v)=v,\;\all{v}\in V$ (projection
operator). Any such projection operator $\alpha$, defines a 1-form
$\widetilde{\alpha}\in\cohom[1]{L}$:
$$
\widetilde{\alpha}(X)=[\alpha(X)],\;\all{X}\in L
$$
\begin{lem}
For any two projections $\alpha,\beta:L\To V$, the forms $\pi(\widetilde{\alpha})$ and
$\pi(\widetilde{\beta})$ in $\cohom[1]{L}_1$, are equal and closed.
\end{lem}
\begin{prf}
For such $\alpha$ and $\beta$ we have: $(\alpha-\beta)(v)=v-v=0,\;\all{v}\in V$, which implies
that $i_v(\widetilde{\alpha}-\widetilde{\beta})=0\;\imply\;\widetilde{\alpha}-\widetilde{\beta}\in\cohom[1]{L/V}\;
\imply\;\pi(\widetilde{\alpha})=\pi(\widetilde{\beta})$.\\
For any $v\in V$ and $X\in L$, we have:
$$
\begin{array}{l}
(d\widetilde{\alpha})(v,X)=v(\widetilde{\alpha}(X))-X(\widetilde{\alpha}(v))-\widetilde{\alpha}([v,X])=\\
\\
=-X([v])-[[v,X]]=-[[X,v]]-[[v,X]]=0
\end{array}
$$
Therefore, we obtain that $i_v(d\widetilde{\alpha})=0,\;\all{v}\in V$, which means that the form
$d\widetilde{\alpha}$ belongs to the submodule $\cohom[2]{L/V}$, or equivalently:
$\pi(d\widetilde{\alpha})=d\pi(\widetilde{\alpha})=0\;\imply\;\pi(\widetilde{\alpha})$ is closed
in $\cohom[1]{L}_1$.
\begin{rem}
$d\widetilde{\alpha}-d\widetilde{\beta}=d(\widetilde{\alpha}-\widetilde{\beta})\;\imply$ the forms
$\widetilde{\alpha}$ and $\widetilde{\beta}$ are cohomological in $\cohom[2]{L/V}$, and their
cohomology class is exactly $\sigma_1(\pi(\widetilde{\alpha}))\in H^2(L/V,\;H_1(V,A_V))$.
\end{rem}
\end{prf}\medskip\\
As it follows from the above lemma, for any projection homomorphism $\alpha~:~L~\To~V$, the element
$\pi(\widetilde{\alpha})$ in $\cohom[1]{L}_1$ is one and the same. Let us denote this element
(and also its cohomology class in $H^1(L,H_V)_1$) by $\Delta$. The element $\sigma_1(\Delta)$ in
$H^2(L/V,\;H_1(V,A_V))$ we call the \emph{characteristic class} of the Lie algebra ideal $V$ in $L$.
\begin{rem}
It is clear that if V is an ideal in the Lie algebra L, then the space
$[V,V]=\set{\jami{i}{}a_i[u_i,v_i]\;\vbar\;a_i\in A_V,\,u_i,v_i\in V}$ is also an ideal in $L$
(an in $V$, too). We can consider the Lie algebra $\widetilde{L}=L/[V,V]$ and its commutative
subalgebra $\widetilde{V}=V/[V,V]$, which is also an ideal in $\widetilde{L}$. It is clear that
the homology space $H_1(V,A_V)$ is the same as $\widetilde{V}$ and the quotient $\widetilde{L}/\widetilde{V}$
is the same as $L/V$. One can consider the following short exact sequence
\begin{equation}\label{seq2}
0\;\To\;\widetilde{V}\;\To\;\widetilde{L}\;\To\;L/V\;\To\;0
\end{equation}
As the Lie algebra $\widetilde{V}$ is commutative, in this situation, for any connection form
$\alpha:\widetilde{L}\To\widetilde{V}$ (a splitting of the short exact sequence \ref{seq2}), its
curvature form coincides with $d\alpha\in\Omega^2_{A_V}(L/V,\widetilde{V})$, and the cohomology
class of the form $d\alpha$ in $d\alpha\in\Omega^2_{A_V}(L/V,\widetilde{V})$ coincides with the
characteristic class of the ideal $V$. Therefore, to study the properties of the characteristic class of
a Lie algebra ideal, we can consider the case when the ideal is commutative.
\end{rem}
\section[The Case of a Fiber Bundle]{\large{The Case of a Fiber Bundle}}
The construction described in the previous section could be used in the case of fiber bundles,
which in the classical case gives the well-known characteristic classes of this fiber bundle.
In this section we consider this case from the algebraic point of view.

\newcommand{\chia}[1]{mul(#1)}

Let $S$ be a module over the algebra $A$. Denote by $Diff^1(S)$ the space of differential operators
on $S$, of order $\leq1$. That is: any $X\in Diff^1(S)$ is a $\Complex$-linear (or $\Real$-linear, if $A$ is an algebra over $\Real$) map $X:S\To S$,
such that, for any two $a,b\in A$, one has $[[X,\chia{a}],\chia{b}]=0$, where $\chia{a}$
(and $\chia{b}$) denotes the ``multiplication by $a$ (and $b$)'' operator on the $A$-module $S$.
From this definition easily follows that for any
$a\in A$, the operator $[X,\chia{a}]:S\To S$ is an element of $End_A(S)$, which is the space of
$A$-module endomorphisms of $S$. Let us denote by $\chia{A}$ the subspace of $End_A(S)$ generated
by the operators of the type $\chia{a},\;a\in A$; and by $trans(S)$ the subspace of $Diff^1(S)$
consisting of such $X\in Diff^1(S)$ that $[X,\chia{a}]\in\chia{A},\;\all{a}\in A$. It is clear
that $trans(S)$ is a Lie algebra.
\begin{rem}
In the case when $S$ is a $C^\infty(M)$-module of smooth sections of a vector bundle over some
smooth manifold $M$, $trans(S)$ is the Lie algebra of the Lie group of automorphisms of this vector
bundle.
\end{rem}
It is easy to verify that for any fixed element $X\in trans(S)$ the mapping
$\chia{A}\ni\chia{a}\mapsto[X,\chia{a}]\in\chia{A}$ is a derivation operator on $\chia{A}$ (i.e., satisfies the Leibnitz's rule).
For simplicity, assume that $A=\chia{A}$ (otherwise, one can reduce $A$ to $\chia{A}$). Let us
denote the map $trans(S)\ni X\mapsto[X,\,\cdot]\in Der(A)$ by $\pi$.
\begin{defn}
We call the $A$-module $S$ a \emph{fiber bundle module} if the Lie algebra homomorphism
$\pi:trans(S)/End_A(S)\To Der(A)$ is an epimorphism.
\end{defn}
As it follows from this definition, for any $X\in trans(S),\;s\in S$ and $a\in A$ we have
$X(as)=\pi(X)(a)\cdot s+a\cdot X(s)$.

One has the following short exact sequence of Lie algebra homomorphisms which is also an exact
sequence of $A$-module homomorphisms:
\begin{equation}\label{Atiyah_exact_sequence}
0\;\To\;End_A(S)\equiv aut(S)\;\To\;trans(S)\;\stackrel{\pi}{\To}\;Der(A)\;\To\;0
\end{equation}
In the classical differential geometry, this short exact sequence is known as the \emph{Atiyah
sequence} (see \cite{Atiyah}).

A connection on the fiber bundle $(S,\pi)$ is a splitting of the short exact sequence \ref{Atiyah_exact_sequence}:
$Der(A)\ni X\mapsto \nabla_X\in trans(S)$. From the definition of the space $trans(S)$, follow the well-known
properties of the connection:
$$
\begin{array}{ll}
&\nabla_{aX}(s)=a\nabla_X(s) \\
\textrm{and}&\\
&\nabla_X(as)=\pi(\nabla_X)(a)\cdot s+a\cdot\nabla_X(s)=X(a)\cdot s+a\cdot\nabla_X(s)
\end{array}
$$
for any $a\in A,\;X\in Der(A)\textrm{ and }s\in S$.

Any such splitting (connection) is equivalent to a homomorphism of $A$-modules
$\alpha:trans(S)\To aut(S)$,
such that $\alpha(U)=U,\;\all{U}\in aut(S)$. The map $X\mapsto\nabla_X$ is a Lie algebra
homomorphism if and only if, the kernel of the map $\alpha$ is a Lie subalgebra in $trans(S)$.
The deviation of $ker(\alpha)$ from being a Lie algebra is a 2-form on $trans(S)$ with values in
$aut(S)$, in the classical differential geometry known as the curvature of the connection
$\alpha$:
$$
[X-\alpha(X),Y-\alpha(Y)]=[X,Y]-[X,\alpha(Y)]-[Y,\alpha(X)]+[\alpha(X),\alpha(Y)]
$$
As $aut(S)$ is a kernel of a Lie algebra homomorphism, it is an ideal in the Lie algebra $trans(S)$,
therefore, we have that
$$[X,\alpha(Y)],\,[Y,\alpha(X)],\,[\alpha(X),\alpha(Y)]\in aut(S)$$
This implies that
$$
\begin{array}{l}
\alpha([X-\alpha(X),Y-\alpha(Y)])=\\
\\
=\alpha([X,Y])-[X,\alpha(Y)]+[Y,\alpha(X)]+[\alpha(X),\alpha(Y)]=\\
\\
=-(d\alpha)(X,Y)+[\alpha(X),\alpha(Y)]
\end{array}
$$
which, itself, implies that the homomorphism
$\alpha:trans(S)\To aut(S)$ induces a splitting of the exact sequence \ref{Atiyah_exact_sequence}
as an exact sequence of Lie algebra homomorphisms iff the form $\omega(X,Y)=(d\alpha)(X,Y)-[\alpha(X),\alpha(Y)]$
is 0.

Applying the construction described in the previous section, to this situation, we have the
following: for a connection form $\alpha:trans(S)\To aut(S)$, which is a homomorphism of $A$-modules,
and defines a splitting of the exact sequence \ref{Atiyah_exact_sequence}, consider the induced
1-form with values in the 1-homology space $\widetilde{\alpha}:trans(S)\To H_1(aut(S))$. Its
differential, $d\widetilde{\alpha}$ is a 2-form on the Lie algebra $trans(S)/aut(S)$ which is
isomorphic to $Der(A)$. The cohomology class of the form $d\widetilde{\alpha}$ in
$\Omega^2_A\pars{Der(A),\,H_1(aut(S))}$ is independent of a choice of the connection $\alpha$.

This approach gives an interesting interpretation of the classical case, where we have a vector
bundle over a smooth manifold. Consider this case in more details.

Let $M$ be a smooth manifold, and $\pi:E\To M$ be a complex vector bundle with Hermitian structure;
$A=C^\infty(M)$ and $S=\Gamma(E)$ be the space of smooth sections of the bundle $\pi:E\To M$
(which is also a module over the algebra $A$). In this situation, let $trans(S)$ be the Lie algebra
of the group of unitary transformations of the fiber bundle $\pi:E\To M$ and $aut(S)$ be the Lie algebra
of the group of unitary automorphisms of this fiber bundle (i.e., such unitary transformations
that are identical on the base of the fiber bundle).

It is clear that $aut(S)$ is the same as the space of sections of the bundle $aut(\pi):aut(E)\To M$,
the fiber of which at a point $x\in M$ is the space of anti-Hermitian operators on the Hermitian
complex vector space $\pi^{-1}(x)$. And the homology space $H_1(aut(S))$ is the same as the
space of the sections of the fiber bundle $\widetilde{\pi}:\widetilde{E}\To M$, the fiber of which
at a point $x\in M$ is the homology space $H_1(u(\pi^{-1}(x)))$. Here $u(\pi^{-1}(x))$ denotes
the Lie algebra of anty-Hermitian operators on the space $\pi^{-1}(x)$. If the fiber of the bundle
$\pi:E\To M$ is finite-dimensional, then the homology space $H_1(u(\pi^{-1}(x)))$ is one-dimensional
and the homology class of the element $i\cdot\textrm{Id}$, where $\textrm{Id}$ is the identical
operator on the space $\pi^{-1}(x)$, gives a canonical basis of $H_1(u(\pi^{-1}(x)))$ as a vector
space over $\mathbb{R}$. Therefore, the homology space $H_1(aut(S))$ is canonically isomorphic
to the algebra $A=C^\infty(M)$ and the characteristic class of the ideal $aut(S)$ in the Lie
algebra $trans(S)$ is the element of the De Rham cohomology space
$H^2\pars{trans(S)/aut(S),\,H_2(aut(S))}\cong H^2\pars{Der(A),A}\cong H^2(M,\mathbb{R})$.
This characteristic class, of course, coincides with the first Chern class of the fiber bundle
$\pi:E\To M$. Other classes are obtained by the characteristic polinomials on $aut(S)$, which,
in fact, are polinomials on $H_1(aut(S))$.
\newcommand{\lder}[1]{\mathcal{L}_{#1}}
\section[Connection Conserving Infinitesimal Transformations]{\large{Connection Conserving Infinitesimal\\Transformations}}
Let a triple $(L,S,A)$, where $L$ is a Lie algebra, $A$ is a commutative, associative algebra over
$\Complex$ or $\Real$ and $S$ is a module over the algebra $A$, is a Lie module (see Definition
\ref{Lie_module_definition}). Let $\alpha$ be a 1-form on $L$ with values in the module $S$, i.e.,
$\alpha\in\Omega^1_A(L,S)$. Denote by $L_\alpha$ the subspace of $L$ consisting of such elements
$X\in L$, that $\lder{X}\alpha=0$, where the operator $\lder{X}$ is an operator of Lie derivation:
$\lder{X}=i_X\circ d+d\circ i_X$. It can be verified by direct calculations that the operators
$\lder{X}$ and $i_Y$, for any $X,Y\in L$ satisfy the conditions $[\lder{X},d]=0$ and
$[\lder{X},i_Y]=i_{[X,Y]}$, from which follows that the mapping $X\mapsto\lder{X}$ is a Lie algebra
homomorphism ($[\lder{X},\lder{Y}]=\lder{[X,Y]}$). This implies that the subspace $L_\alpha$ is a
Lie subalgebra in $L$.
\begin{lem}
The kernel of the restricted mapping $\alpha:L_\alpha\To S$ is a Lie algebra ideal in $L_\alpha$.
\end{lem}
\begin{prf}
For any $V\in ker(\alpha)$ and $X\in L_\alpha$, we have
$$
\begin{array}{l}
\lder{V}\alpha=0\;\imply\;i_V(d\alpha)=0;\\
\\
(\lder{X}\alpha)(V)=0\;\imply\;V\alpha(X)+(d\alpha)(X,V)=0\;\imply\;V\alpha(X)=0;\\
\\
\underbrace{(d\alpha)(X,V)}_0=X\underbrace{\alpha(V)}_0-\underbrace{V\alpha(X)}_0-\alpha([X,V])=0\;\imply\\
\\
\imply\;\alpha([X,V])=0\;\imply\;[X,V]\in ker(\alpha).
\end{array}
$$
\end{prf}\bigskip\\
The above lemma implies that the quotient space $L_\alpha/(ker(\alpha)\cap L_\alpha)$ inherits the Lie algebra
structure from $L_\alpha$. Therefore, we have a Lie algebra structure on $\alpha(L_\alpha)\subset S$.
\begin{lem}\label{lemma13}
For any two elements $X,Y\in L_\alpha$, we have that $\alpha([X,Y])=(d\alpha)(X,Y)$.
\end{lem}
\begin{prf}
The condition $\lder{X}\alpha=\lder{Y}\alpha=0$, implies:
$$
(\lder{X}\alpha)(Y)=Y\alpha(X)+X\alpha(Y)-Y\alpha(X)-\alpha([X,Y])=0
$$
and
$$
(\lder{Y}\alpha)(X)=X\alpha(Y)+Y\alpha(X)-X\alpha(Y)-\alpha([Y,X])=0
$$
consequently
$$
\begin{array}{l}
X\alpha(Y)-\alpha([X,Y])=Y\alpha(X)+\alpha([X,Y])=0\;\imply\\
\\
\imply\;X\alpha(Y)-Y\alpha(X)-\alpha([X,Y])=\alpha([X,Y])\;\imply\\
\\
\imply\;(d\alpha)(X,Y)=\alpha([X,Y])
\end{array}
$$
\end{prf}
\begin{lem}\label{subalg_lemma}
If $S=A$ and there exists such element $\eta\in L_\alpha$ that $\alpha(\eta)=1$ then the subset
$\alpha(L_\alpha)$ is a subalgebra in the algebra $A$.
\end{lem}
\begin{prf}
Under these conditions, any $X\in L_\alpha$ can be represented as $X=a\cdot\eta+X'$, where $a\in A$
and $X'\in ker(\alpha)$: $X'=X-\alpha(X)\cdot\eta$. The equality $\lder{X}\alpha=0$ implies:
$\lder{X'}\alpha=-da$ ($\;\Leftrightarrow\;i_{X'}(d\alpha)=-da$). If $a=\alpha(X)$ and $b=\alpha(Y)$,
then $X=a\cdot\eta+X_a$ and $Y=b\cdot\eta+X_b$, where $X_a,X_b\in ker(\alpha)$ (though the correspondence $a\mapsto X_a$,
generally, is not single-valued mapping). Consider the element $W=ab\cdot\eta+(aX_b+bX_a)$ and
verify that $\lder{W}\alpha=0$. From the property of the Lie derivation:
$\lder{aX}\omega=a\cdot\lder{X}\omega+da\wedge i_X(\omega)$, follows that
$$
\lder{W}\alpha=d(ab)-adb-bda=d(ab)-d(ab)=0
$$
Hence, we obtain that, for the element $W=ab\cdot\eta+(aX_b+bX_a)$: $\alpha(W)=ab$ and $\lder{W}\alpha=0$,
which, itself, implies that if $a,b\in\alpha(L_\alpha)$, then $ab\in\alpha(L_\alpha)$. As $1\in\alpha(L_\alpha)$,
we obtain that $\alpha(L_\alpha)$ is a subalgebra of $A$, with unit.
\end{prf}

Let us denote the bracket corresponding to the induced Lie algebra structure on $\alpha(L_\alpha)$,
by $\{\;,\;\}$.

The condition $\lder{\eta}\alpha=0$ implies $i_\eta(d\alpha)=0$. This fact, together with the
lemma \ref{lemma13}, gives that, for any $a,b\in\alpha(L_\alpha)$, we have that
$\set{a,b}=(d\alpha)(X_a,X_b)$. This equality, together with the lemma \ref{subalg_lemma}, implies
the following
\begin{prop}
If $S=A$ and there exists such element $\eta\in L_\alpha$ that $\alpha(\eta)=1$, then $\alpha(L_\alpha)$
is a subalgebra in the algebra $A$, and the algebra $\alpha(L_\alpha)$ together with the bracket
induced from $L_\alpha$ is a Poisson algebra.
\end{prop}
\begin{prf}
For $a,b,c\in\alpha(L_\alpha)$ we have $X=a\eta+X_a,Y=b\eta+X_b,Z=c\eta+X_c\in L_\alpha$. We have,
also, that the element in $L_\alpha$ corresponding to $ab\in\alpha(L_\alpha)$ is
$W=bc\eta+(bX_c+cX_b)$. therefore:
$$
\begin{array}{l}
\set{a,bc}=d\alpha(X_a,bX_c+cX_b)=b(d\alpha)(X_a,X_c)+c(d\alpha(X_a,X_b))=\\
\\
=b\set{a,c}+\set{a,b}c
\end{array}
$$
\end{prf}\bigskip\\
Further we consider this situation, in more details, for the case of a contact manifold.
\section[The Canonical Bivector and Vector Fields]{\large{The Canonical Bivector and Vector Fields on a Contact Manifold}}
A pair $(M,\alpha)$, where $M$ is a $2n+1$-dimensional smooth manifold and $\alpha$ is a
differential 1-form on it, such that $\alpha_x\wedge(d\alpha)_x^n\neq0$ for all points
$x\in M$ is called a \emph{contact manifold} with contact structure given by the form
$\alpha$.

For any point $x\in M$, let
$$
\begin{array}{ll}
              & \kernel{\alpha}_x=\set{u\in T_xM\,|\,\alpha_x(u)=0}           \\
 \textrm{and} &                                                               \\
              & \kernel{d\alpha}_x=\set{u\in T_xM\,|\,(d\alpha)_x(u,\cdot)=0}
\end{array}
$$
The condition $\alpha_x\wedge(d\alpha)_x^n\neq0$, implies that
$T_xM=\kernel{\alpha}_x\oplus\kernel{d\alpha}_x,\;\forall x\in M$.

\newcommand{\tildom}{\widetilde{\omega}}
\newcommand{\pardif}[1]{\frac{\partial}{\partial#1}}

Denote the differential form $d\alpha$ by $\omega$ and define the homomorphism of fiber
bundles $\tildom:TM\To T^*M$ as
$$
\tildom_x(u)(v)=\omega_x(u,v),\;\forall\,x\in M,\,\forall\,u,v\in T_xM
$$
This homomorphism induces the homomorphism of $C^\infty(M)$-modules from the space of
vector fields on the manifold $M$ to the space of differential 1-forms on $M$. We denote
this homomorphism also by $\tildom:V^1(M)\To\Omega^1(M)$. The latter could be extended to
the tensor fields of higher degrees, and we obtain the homomorphism of the graded
exterior algebras, from the space of antisymmetric covariant tensor fields to the space
of differential forms on the manifold $M$
$$
\wedge\tildom=\oplus_{i=0}^\infty\wedge^i\tildom:%
V(M)=\oplus_{i=0}^\infty V^i(M)\To\Omega(M)=\oplus_{i=0}^\infty\Omega^i(M)
$$
Where $V^i(M),\;i=0,\ldots,\infty$ denotes the space of covariant tensor fields
(multivector fields) of degree $i$.

As $V^1(M)=\kernel{\alpha}\oplus\kernel{\omega}$, the map
$\tildom:\kernel{\alpha}\To\Omega^1(M)$ is a monomorphism. The kernel of the map
$\wedge\tildom$ is the ideal generated by $\kernel{\omega}$, and the map
$$
\wedge\tildom:\wedge\kernel{\alpha}\To\Omega(M)
$$
is also a monomorphism.

According to the Darboux theorem, there exists a local coordinate system
$x_0,x_1,\ldots,x_{2n}$, on the contact manifold $M$, such that the contact form $\alpha$
in this coordinate system is expressed as
$$
\alpha=dx_0+\sum\limits_{i=1}^nx_{2i-1}dx_{2i}
$$
Such a coordinate system is called the \emph{canonical coordinate system}. This implies
that in such coordinate system $\omega=\sum\limits_{i=1}^n\wedge dx_{2i-1}dx_{2i}$ and
the mapping $\tildom:V^1(M)\To\Omega^1(M)$ can be written as
$$
\tildom\pars{\pardif{x_0}}=0,\;\tildom\pars{\pardif{x_{2i-1}}}=%
dx_{2i},\;\tildom\pars{\pardif{x_{2i}}}=-dx_{2i-1}\quad i=1,\ldots,n.
$$
\begin{lem}
The differential form $\omega$ is an element of the space $\image{\wedge^2\tildom}$, for
the mapping $\wedge^2\tildom:V^2(M)\To\Omega^2(M)$.
\end{lem}
\begin{prf}
From the representations of the form $\omega$ and the mapping $\tildom$ in the canonical
coordinate system, easily follows that the map $\wedge^2\tildom:V^2(M)\To\Omega^2(M)$,
carries the bivector field $w=\sum\limits_{i=1}^n\pardif{x_{2i-1}}\wedge\pardif{x_{2i}}$
into the form $\omega$:
$$
\tildom(w)=\sum\limits_{i=1}^n\tildom\pars{\pardif{x_{2i-1}}\wedge\pardif{x_{2i}}}=%
\sum\limits_{i=1}^ndx_{2i}\wedge(-dx_{2i-1})=\sum\limits_{i=1}^ndx_{2i-1}\wedge dx_{2i}%
$$
\end{prf}

It is clear that
$V^2(M)=\bigl(\kernel{\omega}\wedge V^1(M)\bigr)\oplus\bigl(\wedge^2\kernel{\alpha}\bigr)$, %
and $\kernel{\wedge^2\omega}=\kernel{\omega}\wedge V^1(M)$. Therefore, the projection of
the bivector field $w$ on the subspace $\wedge^2\kernel{\alpha}$ is also carried in the
form $\omega$, by the map $\wedge^2\tildom$. Let us denote this projection by $\mu$
(though the bivector field $w$ could be only local, its projection is a global bivector
field). The representation of $\mu$ in the canonical coordinate system is
\begin{equation}\label{mu_coordinate_formula}
\mu=\sum\limits_{i=1}^n\pardif{x_{2i-1}}\wedge\pars{\pardif{x_{2i}}-x_{2i-1}\pardif{x_0}}
\end{equation}
To summarize, we can state that there exists a unique, canonical bivector field
$\mu\in\wedge^2\kernel{\alpha}$ such that $\wedge^2\tildom(\mu)=\omega$. We call the
bivector field $\mu$, the \emph{canonical bivector field} on the contact manifold
$(M,\alpha)$.

Any bivector field, on a smooth manifold, defines a bracket on the algebra of the smooth
functions on that manifold, so does $\mu$
$$
\{f,g\}=(df\wedge dg)(\mu),\quad\forall\,f,g\in C^\infty(M)
$$
The same bracket can be uniquely defined by its property (see \cite{Hurt}):
$$
\{f,g\}\alpha\wedge\omega^n=ndf\wedge dg\wedge\alpha\wedge\omega^{n-1}
$$
It is clear that such bracket is antisymmetric and biderivative, but to satisfy the
Jacoby identity, the bivector field $\mu$ needs to be involutive, i.e. satisfy the
condition $[\mu,\mu]=0$ (see \cite{Lichnerowicz}). Here, the bracket $[\cdot\;,\;\cdot]$
denotes the supercommutator operation on the Lie superalgebra of antisymmetric covariant
tensor fields. The coordinate representation of the bivector field $\mu$ (see the formula
\ref{mu_coordinate_formula}), implies that $[\mu,\mu]=\pardif{x_0}\wedge W$, where $W$ is
some bivector field; therefore, the bracket $\{\cdot\;,\;\cdot\}$ cannot satisfy the
Jacoby identity on the entire algebra $C^\infty(M)$.

\newcommand{\basef}{\mathcal{F}_B(M)}
\newcommand{\basev}{V^1_I(M)}

For any bivector field $\mu$ and any triplet of functions $f,g,h\in C^\infty(M)$, we have
the following equality for the bracket defined by $\mu$ (see \cite{Me}):
$$
\{\{f,g\},h\}+\{\{g,h\},f\}+\{\{h,f\},g\}=\frac{1}{2}(df\wedge dg\wedge dh)([\mu,\mu])
$$
As the supercommutator $[\mu,\mu]$ is an element of the ideal generated by
$\kernel{\omega}$, we have that the following subalgebra
$$
\basef=\set{\varphi\in C^\infty(M)\,|\,X(\varphi)=0,\;\forall\,X\in\kernel{\omega}}
$$
is the subalgebra (maximal) of the commutative algebra $C^\infty(M)$, which is closed
under the bracket defined by the canonical bivector field $\mu$ and on which the Jacoby
identity is satisfied. That is, the algebra $\basef$ is a Poisson algebra under the
bracket defined by the canonical bivector field.

Further, we call the elements of the submodule $\kernel{\alpha}$, the \emph{horizontal}
vector fields and the elements of the submodule $\kernel{\omega}$, the \emph{vertical}
vector fields. The algebra $\basef$ is called the algebra of \emph{basic functions} on the
contact manifold $(M,\alpha)$ (see \cite{Hurt}).

For any $w\in V^m(M)$ and $\beta\in\Omega^n(M)$, where $m\geq n$, we denote by
$\widetilde{w}(\beta)$ the element of the space $V^{m-n}(M)$ such that for any
$\lambda\in\Omega^{m-n}(M):\quad\lambda(\widetilde{w}(\beta))=(\beta\wedge\lambda)(w)$.

For the canonical bivector field $\mu$, we have that $[\mu,\mu]\in V^3(M)$, therefore:
$\widetilde{[\mu,\mu]}(\omega)\in V^1(M)$. Let us denote the vector field
$-\widetilde{[\mu,\mu]}(\omega)\in V^1(M)$ by $\eta$. Find the representation of this
vector field in the canonical coordinate system:
$$
\begin{array}{l}
[\mu,\mu]=-\left[\sum\pardif{x_{2i-1}}\wedge\pardif{x_{2i}},-\sum x_{2i-1}\pardif{x_{2i-1}}\wedge\pardif{x_0}\right]=\\%
\\
=\sum\pardif{x_{2i}}\wedge\pardif{x_{2i-1}}\wedge\pardif{x_0}\quad\Rightarrow\quad\widetilde{[\mu,\mu]}(\omega)=-\pardif{x_0}\quad\Rightarrow\quad\eta=\pardif{x_0}%
\end{array}
$$
The vector field $\eta$ can be uniquely characterized by the properties:
$i_\eta\alpha=1,\;i_\eta\omega=0$, and is known as the \emph{canonical} vector field on
the contact manifold $(M,\alpha)$ (see \cite{Hurt}).

It is clear that the module $\kernel{\omega}$ is one-dimensional, therefore, the vector
field $\eta$ can be considered as its basis, after which the Poisson algebra of the basic
functions, $\basef$, can be defined as $\basef=\kernel{\eta}$, where $\eta:C^\infty(M)\To
C^\infty(M)$ is a derivation operator.
\newcommand{\ham}[1]{\widetilde{\mu}(#1)}
\newcommand{\smooth}[1]{C^\infty(#1)}
\newcommand{\vf}[2]{V^#1(#2)}
\newcommand{\pardiff}[2]{\frac{\partial#1}{\partial#2}}
\section[Properties of Basic Functions and Invariant Vector Fields]{\large{Some Properties of Basic Functions and Invariant Vector Fields}}
For any $f\in C^\infty(M)$, we can consider the vector field $\ham{df}$. Let us formulate
and verify some properties of the vector fields of the type $\ham{df},\;f\in\smooth{M}$.
\begin{lem}\label{ham_is_horizontal_lemma}
For any $f\in\smooth{M}$, the vector field $\ham{df}$ is horizontal.
\end{lem}
\begin{prf}
By definition of the operator $\widetilde{\mu}:\Omega^1(M)\To V^1(M)$, we have the
following: $\alpha(\ham{df})=(df\wedge\alpha)(\mu)$, but $\mu$ is an element of
$\wedge^2\kernel{\alpha}$, which implies that $(df\wedge\alpha)(\mu)=0$.
\end{prf}

\begin{prop}
The equality $df=-\omega(\ham{df},\,\cdot)$ is true if and only if $f$ is a base
function.
\end{prop}
\begin{prf}
If for some vector field $X$ we have that $df=-\omega(X,\,\cdot)$, then
$\eta(f)=-\omega(X,\eta)=0$. Therefore, in this case $f\in\basef$. Now, the task is to
verify that if $f\in\basef$ then $df=-\omega(\ham{df},\,\cdot)$. Using the canonical
coordinate system, we obtain:
$$
\begin{array}{l}
f\in\basef\quad\Rightarrow\quad\pardiff{f}{x_0}=0\quad\Rightarrow\\
\\
\ham{df}=\sum\pars{\pardiff{f}{x_{2i-1}}\pardif{x_{2i}}-\pardiff{f}{x_{2i}}\pardif{x_{2i-1}}}-\sum x_{2i-1}\pardiff{f}{x_{2i-1}}\pardif{x_0}\quad\Rightarrow\\%
\\
\omega(\ham{df},\,\cdot)=-\pardiff{f}{x_{2i}}dx_{2i}-\pardiff{f}{x_{2i-1}}dx_{2i-1}=-df
\end{array}
$$
\end{prf}

To summarize, we can state that: for any function $f\in C^\infty(M)$, the equality
$df=-\omega(X,\,\cdot)$ for some vector field $X$ is true if and only if $f\in\basef$;
and as the form $\omega$ is non-degenerated on the submodule of horizontal vector fields,
in this case, there exists one and only one such horizontal vector field $X$, which equal
to $\ham{df}$.
\begin{cor}
For the Poisson algebra $\basef$ the following equality is true
$\{f,g\}=\omega(\ham{df},\ham{dg})$.
\end{cor}
Let us denote by $\basev$ the set of vector fields on the manifold $M$ commuting with
$\eta:\;\basev=\set{X\in\vf{1}{M}\,|\,[\eta,X]=0}$, and call them the \emph{invariant
vector fields} (see \cite{Michel1}). It is clear that $\basev$ is a Lie subalgebra in
$\vf{1}{M}$ and a module over the algebra $\basef$.
\begin{lem}
For any $f\in\basef$ the horizontal vector field $\ham{df}$ is an element of the space
$\basev$.
\end{lem}
\begin{prf}
The proof easily follows from the coordinate representation of the vector field
$\ham{df}$:
$$
\ham{df}=\sum\pars{\pardiff{f}{x_{2i-1}}\pardif{x_{2i}}-\pardiff{f}{x_{2i}}\pardif{x_{2i-1}}}-
\sum x_{2i-1}\pardiff{f}{x_{2i-1}}\pardif{x_0}
$$
\end{prf}

The subalgebra $\basef$ in $\smooth{M}$ is invariant under the action of the elements of
the space $\basev$:
$$
\begin{array}{l}
\varphi\in\basef,\;X\in\basev\;\Rightarrow\;\eta(X(\varphi))=[\eta,X](\varphi)+%
X(\eta(\varphi))=0\;\Rightarrow\\
\\
\Rightarrow\quad X(\varphi)\in\basef
\end{array}
$$
Hence, we have a natural homomorphism of Lie algebras
$$\pi:\basev\To Der(\basef)$$
where $Der(A)$ denotes the Lie algebra of derivations for any algebra $A$:
$$
\begin{array}{lll}
Der(A)&=&\{X:A\To A\,|\,X\textrm{ is linear, and }\\
&&\\
&&X(ab)=X(a)b+aX(b),\;\,\forall\,a,b\in A\}
\end{array}
$$
The kernel of the homomorphism $\pi$ is the subalgebra
$$\basef\cdot\eta=\set{\varphi\cdot\eta\,|\,\varphi\in\basef}$$
Consider the following short exact sequence
\begin{equation}\label{exact_sequence_of_basefields}
0\;\To\;\basef\cdot\eta\;\hookrightarrow\;\basev\;\stackrel{\pi}{\To}\;\image{\pi}\;\To\;0
\end{equation}
The submodule $\basev\bigcap\kernel{\alpha}$ gives a splitting of the short exact
sequence \ref{exact_sequence_of_basefields}, but this splitting is not a Lie algebra
homomorphism, because, for any two horizontal vector fields $X$ and $Y$, we have
$\alpha([X,Y])=-\omega(X,Y)$. Any splitting of the short exact sequence
\ref{exact_sequence_of_basefields} is equivalent to a choice of a $\basef$-submodule
$\mathcal{H}$ of the module $\basev$. This choice, itself, is equivalent to a 1-form
$\beta$ on $\basev$ with values in $\basef$. The form $\beta$, actually, is the
projection operator on the submodule $\basef\cdot\eta$, corresponding to the
decomposition $\basev=(\basef\cdot\eta)\oplus\mathcal{H}$. The form $\beta$ can be
characterized by the following properties: $i_\eta\beta=1$ and $L_\eta(\beta)=0$, where
$L_\eta$ denotes the Lie derivation: $L_\eta=d\circ i_\eta+i_\eta\circ d$ (for
$X\in\basev$:
$(L_\eta\beta)(X)=(d\beta)(\eta,X)=\eta\underbrace{\beta(X)}_{\in\basef}-%
X\underbrace{\beta(\eta)}_1-\beta(\underbrace{[\eta,X]}_0)=0$). For each point $x\in M$,
the subalgebra $\basev$ gives the entire tangent space at this point, which implies that
the form $\beta:\basev\To\basef$ could be extended to a differential 1-form on the
manifold $M$.

\newcommand{\equco}{\Omega_B(M)}

Let us denote by $\equco$ the subspace of the space $\Omega(M)$ consisting of such
differential forms $\theta\in\Omega(M)$ that $i_\eta\theta=L_\eta(\theta)=0$. Such forms
are called the \emph{basic forms} (see \cite{Michel1}).
For any basic form $\theta$, we have: $L_\eta(d\theta)=dL_\eta(\theta)=0$ and $i_\eta d\theta=L_\eta(\theta)-di_\eta\theta=0$; %
therefore, the subspace $\equco$ is invariant under the action of the differential $d$.
Hence, we can talk about subcomplex $(\equco,d)$ of the De Rham complex $(\Omega,d)$. It
is clear that $\equco$ is closed under the exterior multiplication. Let us denote the
cohomology algebra of the complex $(\equco,d)$ by $H_B(M)$ (basic cohomology, see \cite{Michel1}).

Any differential 1-form $\beta$ on the manifold $M$, corresponding to some splitting of
the short exact sequence \ref{exact_sequence_of_basefields} (i.e., with properties
$\beta(\eta)=1$ and $L_\eta(\beta)=0$) defines a cohomology class in $H_B^2(M)$
corresponding to the differential form $\theta=d\beta$. Any two such forms $\beta_1$ and
$\beta_2$ define one and the same cohomology class in $H_B^2(M)$:
$d\beta_1-d\beta_2=d(\beta_1-\beta_1)$, $\beta_1-\beta_1\in\Omega^1_\eta(M)$.
\begin{prop}\label{split_proposition}
The exact sequence \ref{exact_sequence_of_basefields} of $\basef$-module homomorphisms
can be split as the exact sequence of Lie algebra homomorphisms, if and only if the
cohomology class of the form $\omega=d\alpha$ in $H_\eta^2(M)$ is trivial.
\end{prop}
\begin{prf}
If the splitting defined by some form $\beta$ is such that the submodule $\kernel{\beta}$
is a Lie subalgebra, then for any $X,Y\in\kernel{\beta}$, we have that $(d\beta)(X,Y)=0$,
which, together with the fact that $\basev=(\basef\cdot\eta)\oplus\kernel{\beta}$ and the
properties of the form $\beta$, implies that $d\beta=0$. Consider the differential form
$\gamma=\alpha-\beta$. It is clear that $\gamma\in\Omega_B^1(M)$ and
$d\gamma=d\alpha=\omega$.
\end{prf}
\newcommand{\basemulti}[1][]{V_I^{#1}(M)}
\section[Poisson Cohomologies of Basic Multivector Fields]{\large{Poisson Cohomologies of Basic Multivector Fields}}
For any integer $n\geq0$, let us denote by $\basemulti[n]$ the subspace of the space
$V^n(M)$, consisting of such elements $w$ that $[\eta,w]=0$, where the bracket
$[\,\cdot\;,\;\cdot\,]$ denotes the supercommutator on the Lie superalgebra $V(M)$. Let
us denote the graded space $\bigoplus\limits_{i=0}^\infty\basemulti[i]$ by $\basemulti$.
We call the elements of the space $\basemulti$ the \emph{invariant multivector fields}.
The space $\basemulti$ is a module over the algebra $\basef$, which follows from the
properties of the Schouten bracket: $[\eta,\varphi w]=\eta(\varphi)w+\varphi[\eta,w]=0$
for $w\in\basemulti$ and $\varphi\in\basef$. From the property
$(-1)^{|u||w|}[[u,v],w]+(-1)^{|v||u|}[[v,w],u]+(-1)^{|w||v|}[[w,u],v]=0$, follows that
the space $\basemulti$ is closed under the operation of the supercommutator; and from the
property $[u,v\wedge w]=[u,v]\wedge w+(-1)^{(|u|+1)|v|}v\wedge[u,w]$ follows that it is
closed under the exterior product. To summarize, we can state that $\basemulti$ inherits
the exterior algebra and superalgebra structures from $V(M)$.

The canonical bivector field $\mu$ is an element of $\basemulti[2]$, because
$[\eta,\mu]=0$, which can be easily verified by using of the representations of $\eta$
and $\mu$ in the canonical coordinate system.
\begin{lem}
The ideal generated by the canonical vector field $\eta\in\basemulti[1]$ in the exterior
algebra $\basemulti$ is also an ideal under the supercommutator.
\end{lem}
\begin{prf}
For any $v,w\in\basemulti$, we have:
$$
[w,\eta\wedge v]=\underbrace{[w,\eta]}_0\wedge v+(-1)^{|w|+1}\eta\wedge[w,v]=(-1)^{|w|+1}\eta\wedge[w,v]\in\eta\wedge\basemulti%
$$
\end{prf}

\newcommand{\oper}{\delta_\mu}

The canonical bivector field $\mu$ defines an operator of degree +1:
$$
\oper:V(M)\To V(M),\quad\oper(v)=[\mu,v]
$$
This is the well-known coboundary operator in the case of Poisson manifolds; and it is
well-known also that the property $\oper\circ\oper=0$ is equivalent to $[\mu,\mu]=0$. But
in this case $[\mu,\mu]\neq0$, which implies that the operator $\oper$ is \textbf{not}
coboundary on $V(M)$.
\begin{lem}
The subalgebra $\basemulti$ in $V(M)$ is invariant under the action of the operator
$\oper$.
\end{lem}
\begin{prf}
For any $w\in\basemulti$, we have the following
$$
[\eta,\oper(w)]=[\eta,[\mu,w]]=\pm[\underbrace{[\eta,\mu]}_0,w]\pm[\mu,\underbrace{[\eta,w]}_0]=0%
$$
which implies that $\oper(w)\in\basemulti$
\end{prf}
\begin{lem}
The ideal generated by the canonical vector field $\eta\wedge\basemulti$ is invariant
under the action of the operator $\oper$.
\end{lem}
\begin{prf}
Actually, the statement of this lemma is tautological, because as it was mentioned
$\mu\in\basemulti[2]$ and the ideal $\eta\wedge\basemulti$ is an ideal also under the
supercommutator. Therefore,
$$
\oper(\eta\wedge\basemulti)=[\mu,\eta\wedge\basemulti]\subset\eta\wedge\basemulti
$$
\end{prf}
\begin{lem}
The multivector field $[\mu,\mu]$ is an element of the ideal $\eta\wedge\basemulti$.
\end{lem}
\begin{prf}
This fact can be easily verified by using the representation of $[\mu,\mu]$ in the
canonical coordinate system (see the formula \ref{mu_coordinate_formula}):
$$
[\mu,\mu]=\sum\pars{\pardif{x_{2i}}\wedge\pardif{x_{2i-1}}}\wedge\pardif{x_0}
$$
\end{prf}

As the submodule $\eta\wedge\basemulti$ is a superalgebra ideal in $\basemulti$, we have
that the quotient $\basemulti/(\eta\wedge\basemulti)$ is also a superalgebra under the
supercommutator induced from $\basemulti$. As $[\mu,\mu]\in\eta\wedge\basemulti$, we have
that the induced operator on the quotient algebras
$$
\begin{array}{c}
\left[\oper\right]:\basemulti/(\eta\wedge\basemulti)\To\basemulti/(\eta\wedge\basemulti)\\
\\
\left[\oper\right](p(w))=p([\mu,w])
\end{array}
$$
where $p(w)$ denotes the equivalency class of the element $w\in\basemulti$ in the
quotient $\basemulti/(\eta\wedge\basemulti)$, is a \textbf{coboundary} operator.

\newcommand{\invarvec}[2][]{V^{#1}_I(#2)}
\newcommand{\basevec}[2][]{V^{#1}_B(#2)}
\newcommand{\baseform}[2][]{\Omega^{#1}_B(#2)}

For simplicity, we denote the operator $[\oper]$ by $\oper$ and the quotient algebra
$\invarvec{M}/(\eta\wedge\invarvec{M})$ by $\basevec{M}$; and call its elements the
\emph{basic multivector fields}. So, the algebra of the basic multivector fields,
together with the operator $\oper$ is a graded differential algebra.

The definition of the graded exterior algebras $\baseform{M}$ and $\basevec{M}$, depend
only on the canonical vector field $\eta$, and not on the contact form $\alpha$; but the
definition of the operator $\oper:\basevec{M}\To\basevec{M}$ depends not only on the
vector field $\eta$ but on some bivector field $\mu$ (which is defined by the contact
form), with the properties: $[\eta,\mu]\in I_\eta$ and $[\mu,\mu]\in I_\eta$, where
$I_\eta$ is the ideal in the exterior algebra $\basevec{M}$ generated by the element
$\eta$.

The representation of any basic form $\theta$ on the contact manifold $M$, in the
canonical local coordinate system is:
$$
\theta=\sum\limits_{i_p\neq0,\forall p\in\set{1,\ldots,k}}\varphi_{i_1,\ldots,i_k}dx_{i_1}\wedge\cdots\wedge dx_{i_k}%
,\quad\textrm{ where }\quad\pardiff{\varphi_{i_1,\ldots,i_k}}{x_0}=0
$$
and the representation of any invariant multivector field $w$ in this coordinate system
is:
$$
w=\sum\psi_{j_1,\ldots,j_l}\pardif{x_{j_1}}\wedge\cdots\wedge\pardif{x_{i_l}}%
,\quad\textrm{ where }\quad\pardiff{\psi_{j_1,\ldots,j_l}}{x_0}=0
$$
Take into consideration this representations, the space $\basevec{M}$, locally, could be
identified with the set of such $w$ that
$$
w=\sum\limits_{j_p\neq0,\forall p\in\set{1,\ldots,l}}\psi_{j_1,\ldots,j_l}\pardif{x_{j_1}}\wedge,\cdots,\wedge\pardif{x_{i_l}}%
,\quad\pardiff{\psi_{j_1,\ldots,j_l}}{x_0}=0
$$
From these easily follows that the restriction of the map
$$
\wedge\tildom:V(M)\To\Omega(M)
$$
on the subalgebra of invariant multivector fields, takes its values in the subalgebra of
the basic forms $\baseform{M}$.

\newcommand{\tildmu}{\widetilde{\mu}}

The kernel of the map $\wedge\tildom:\invarvec{M}\To\baseform{M}$ is the ideal generated
by the canonical vector field $\eta$.
\begin{prop}\label{complexes_isomorphism_proposition}
The homomorphism of the exterior superalgebras
$$
\wedge\tildom:\basevec{M}=\invarvec{M}/(\eta\wedge\invarvec{M})\To\baseform{M}
$$
is an isomorphism of the differential complexes $(\basevec{M},\oper)$ and
$(\baseform{M},d)$.
\end{prop}
\begin{prf}
By using of the canonical local coordinates, it is easy to verify that
$d\circ(\wedge\tildom)=(\wedge\tildom)\circ\oper$, and the inverse map is
$$
p\circ(\wedge\tildmu):\baseform{M}\To\basevec{M}
$$
where $p:\invarvec{M}\To\basevec{M}$ is the quotient map and $\tildmu:\Omega^1(M)\To V^1(M)$ %
is defined as:
$\theta(\tildmu(\gamma))=(\gamma\wedge\theta)(\mu),\;\theta,\gamma\in\Omega^1(M)$.
\end{prf}

The isomorphism $\wedge\tildom$ carries the equivalency class of the bivector field $\mu$
in the differential form $\omega=d\alpha$, therefore the proposition
\ref{complexes_isomorphism_proposition}, together with the proposition
\ref{split_proposition}, implies that the exact sequence
\ref{exact_sequence_of_basefields} can be split by a homomorphism of $\basef$-modules and
Lie algebras if and only if the cohomology class of $\mu$ in $\basevec[2]{M}$ is trivial;
i.e., there exists such vector field $V\in\invarvec[1]{M}$ that
$([\mu,V]-\mu)\in\eta\wedge\invarvec{M}$.
\section[Geometric Quantization of Basic Functions]{\large{Geometric Quantization of the Poisson Algebra of Basic Functions}}
The contact form $\alpha$ on the contact manifold $(M,\alpha)$, defines a mapping from the
Lie algebra of the invariant vector fields to the Poisson algebra of basic functions:
$$
\alpha:\invarvec[1]{M}\To\basef
$$
This mapping is surjective, because $\basef\cdot\eta$ is a subalgebra of
$\invarvec[1]{M}$, and on this subalgebra $\alpha(\varphi\cdot\eta)=\varphi$, for each
$\varphi\in\basef$. But $\invarvec[1]{M}\stackrel{\alpha}{\To}\basef$ is not a Lie algebra
homomorphism, because the subalgebra $\basef\cdot\eta$ is commutative Lie subalgebra in
$\invarvec[1]{M}$, while the Lie algebra $(\basef, \{\;,\;\})$, generally, is not commutative.

Hence, we have the following short exact sequence of homomorphisms of $\basef$-modules:
\begin{equation}\label{short_exact_sequence2}
0\;\To\;\kernel{\alpha}_I\;\hookrightarrow\;\invarvec[1]{M}\;\stackrel{\alpha}{\To}\;\basef\;\To\;0
\end{equation}
where $\kernel{\alpha}_I$ denotes the submodule of invariant, horizontal vector fields:
$\invarvec[1]{M}\bigcap\kernel{\alpha}$. Now consider some simple, general situstion. Let
$A$, $B$ and $C$ be linear spaces, and
$$
0\;\To\;A\;\hookrightarrow\;B\;\stackrel{\pi}{\To}\;C\;\To\;0
$$
be a short exact sequence of linear maps, and let $s_0:C\To B$ be some splitting of this exact
sequence. In this situation, there is a one-to-one correspondence between the set of splittings
of the above short exact sequence and the set of linear maps from $C$ to $A$: if
$\varphi:C\To A$ is a linear map, then the map $s=s_0+\varphi:C\To B$ is a splitting, and vice
versa, if $s:C\To B$ is a splitting, then $\varphi=s-s_0$ is a linear map from $C$ to $A$.

In our case, we have two maps: $f\mapsto\ham{df}$ from $\basef$ to $\kernel{\alpha}_I$ and
$f\mapsto f\cdot\eta$ from $\basef$ to $\invarvec[1]{M}$. The secod one is a splitting of the
short exact sequence \ref{short_exact_sequence2}. From these two maps, as it was discussed,
can be constructed a new splitting of the short exact sequence \ref{short_exact_sequence2}:
$$
\basef\ni f\mapsto\widehat{f}=f\cdot\eta+\ham{df}\in\invarvec[1]{M}
$$
\begin{prop}\label{basefunctions_to_contact_automorphisms}
The mapping $f\mapsto\widehat{f}$ is a Lie algebra monomorphism from the Poisson algebra of
basic functions, $\basef$, to the Lie algebra of invariant vector fields $\invarvec[1]{M}$, and
the image of this mapping coincides with the subalgebra of such vector fields $X$ in
$\invarvec[1]{M}$ that $L_X(\alpha)=0$.
\end{prop}
\begin{prf}
It is clear that the set $\set{X\in\invarvec[1]{M}\;|\;L_X(\alpha)=0}$ is a Lie subalgebra in
the Lie algebra $\invarvec[1]{M}$. If $X=f\cdot\eta+\ham{df}$ for some basic function $f$, then
we have $L_X(\alpha)=df+(d\alpha)(\ham{df},\cdot)=0$ which follows from the equality
$df=-(d\alpha)(\ham,\cdot)$, for the basic functions.

If $X\in\invarvec[1]{M}$, then $X$ can be represented as $X=f\cdot\eta+X'$, where $X'$ is a
horizontal vector field. The equality $L_X(\alpha)=0$ implies $df+(d\alpha)(X',\cdot)=0$, but
as it was shown, for any basic function $f\in\basef$, there is one and only one horizontal
vector field $X'$ such that $df=-(d\alpha)(X',\cdot)$, and this vector field is $\ham{df}$.
Therefore, we obtain that $X=f\cdot\eta+\ham{df}$.
\end{prf}

We can consider the map $X\mapsto\alpha(X)$, not only on the invariant vector fields but extend
it to the Lie algebra of all vector fields on the contact manifold $M$:
$V^1(M)\stackrel{\alpha}{\To}\smooth{M}$. This map is a surjective too:
$\alpha(\varphi\eta)=\varphi,\;\forall\,\varphi\in\smooth{M}$. Hence, we can consider the short
exact sequence
\begin{equation}\label{short_exact_sequence3}
0\;\To\;\kernel{\alpha}\;\hookrightarrow\;V^1(M)\;\stackrel{\alpha}{\To}\;\smooth{M}\;\To\;0
\end{equation}
In this case, we have not defined a Lie algebra structure on the commutative algebra $\smooth{M}$
(the bivector field $\mu$ gives such structure only on the subalgebra of the basic functions), but
the mapping $f\mapsto\widetilde{f}=f\cdot\eta+\ham{df}$ can be extended to the entire $\smooth{M}$.
To describe the image of this mapping, let us review some notions from the theory of contact
manifolds.
\begin{defn}[Infinitesimal Contact Transformation]
A vector field $X$ on the contact manifold $(M,\alpha)$ is called an infinitesimal contact
transformation, if exists such function $\varphi\in\smooth{M}$, that $L_X(\alpha)=\varphi\cdot\alpha$
(see \cite{Hurt}).
\end{defn}
Let us denote the set of infinitesimal contact transformations by $cont(M,\alpha)$.
\begin{defn}[Infinitesimal Contact Automorphism]
A vector field $X$ on the contact manifold $(M,\alpha)$ is called an infinitesimal contact
automorphism, if $L_X(\alpha)=0$ (see \cite{Hurt}).
\end{defn}
Let us denote the set of infinitesimal contact automorphisms by $cont_0(M,\alpha)$.

\newcommand{\ctt}[2]{\ensuremath{cont(#1,#2)}}
\newcommand{\cta}[2]{\ensuremath{cont_0(#1,#2)}}

It is clear, that \ctt{M}{\alpha} is a Lie algebra and \cta{M}{\alpha} is its subalgebra. The
Proposition \ref{basefunctions_to_contact_automorphisms} states that \cta{M}{\alpha} is is
isomorphic to the Lie algebra of the basic functions, via the mapping
$$
\basef\ni f\mapsto\widehat{f}=f\cdot\eta+\ham{df}\in\cta{M}{\alpha}
$$
\begin{prop}
The mapping $f\mapsto\widehat{f}=f\cdot\eta+\ham{df}$, is a bijection between the set of
\textbf{all} smooth functions on the contact manifold $M$ and the set \ctt{M}{\alpha}.
\end{prop}
\begin{prf}
Let $X=\widehat{f}$, for some $f\in\smooth{M}$. Consider the form $L_X(\alpha)$. As the vector
field $\ham{df}$ is always horizontal (see Lemma \ref{ham_is_horizontal_lemma}), we have that
$\alpha(\ham{df})=0$, and $L_X(\alpha)=df+\omega(\ham{df},\cdot)$. In the canonical local
coordinate system we have the following
$$
\begin{array}{ll}
    & df=\pardiff{f}{x_0}dx_0+\sum\pars{\pardiff{f}{x_{2i-1}}dx_{2i-1}+\pardiff{f}{x_{2i}}dx_{2i}}\quad\Rightarrow \\
 &   \\
    & \ham{df}=\sum\pars{x_{2i-1}\pardiff{f}{x_0}\pardif{x_{2i-1}}+\pardiff{f}{x_{2i-1}}\pardif{x_{2i}}-\pardiff{f}{x_{2i}}\pardif{x_{2i-1}}-x_{2i-1}\pardiff{f}{x_{2i-1}}\pardif{x_0}} \\
 &   \\
 \Rightarrow & \omega(\ham{df},\cdot)=\sum x_{2i-1}\pardiff{f}{x_0}dx_{2i}-\sum\pars{\pardiff{f}{x_{2i-1}}dx_{2i-1}+\pardiff{f}{x_{2i}}dx_{2i}} \\
 &   \\
 \Rightarrow & df+\omega(\ham{df},\cdot)=\pardiff{f}{x_0}\pars{dx_0+\sum x_{2i-1}dx_{2i}}=\eta(f)\cdot\alpha
\end{array}
$$
which implies that any vector field of the type $\widehat{f}=f\cdot\eta+\ham{df}$ is an element
of the Lie algebra \ctt{M}{\alpha}.

Now, let $X$ be any vector field from \ctt{M}{\alpha}. The vector field $X$ can be decomposed as
the sum of its vertical and horizontal components: $X=f\cdot\eta+W$. The condition
$L_X(\alpha)=\varphi\cdot\alpha$, implies: $df+\omega(W,\cdot)=\varphi\cdot\alpha$. But it was
shown that $df+\omega(\ham{df},\cdot)=\eta(f)\cdot\alpha$. Therefore,
$\omega(W-\ham{df},\cdot)=(\varphi-\eta(f))\cdot\alpha$. Putting $\eta$ in the both sides of this
equality gives
$$
\varphi-\eta(f)=0\;\Rightarrow\;\omega(W-\ham{df},\cdot)=0
$$
The vector fields $W$ and $\ham{df}$ are horizontal, therefore, such is their difference. As the
form $\omega$ is non-degenerated on the module of horizontal vector fields, we obtain that
$W=\ham{df}$.
\end{prf}

So, we obtained that the elements of the Lie algebra \ctt{M}{\alpha} are the vector fields of the
type $X=f\cdot\eta+\ham{df}$, where $f\in\smooth{M}$; and the function $\varphi\in\smooth{M}$,
for which $L_X(\alpha)=\varphi\cdot\alpha$ is the function $\eta(f)$ (or, which is the same:
$\eta(\alpha(X))$). In the case of the basic function, we have that $\eta(f)=0$, and consequently:
$L_{\widehat{f}}(\alpha)=0$.

\newcommand{\scalar}[2]{\langle#1,#2\rangle}
\newcommand{\hilb}{\mathcal{H}}

The homomorphism of Lie algebras $\pi:\invarvec[1]{M}\To Der(\basef)$ can be restricted to the
subalgebra of infinitesimal contact automorphisms. We call the image
$\pi(\cta{M}{\alpha})\subset Der(\basef)$ the \emph{dynamical vector fields}, and denote this
Lie aubalgebra under $dyn(M,\alpha)$. For any two basic functions $f$ and $\varphi$, we have that
$\widehat{f}(\varphi)=(f\cdot\eta+\ham{df})(\varphi)=\ham{df}(\varphi)=\omega(\ham{d\varphi},\ham{df})$.
As the differential 2-form $\omega$ is non-degenerated on the space of horizontal vector fields,
and for each point $x\in M$, the subspace of the tangent space $T_xM$, generated by the set vector
fields of the type $\ham{d\varphi},\;\varphi\in\basef$, coincides with the entire horizontal subspace,
we obtain that
$$
(\widehat{f}(\varphi)=0,\;\forall\varphi\in\basef)\;\Leftrightarrow\;
\ham{df}=0\;\Leftrightarrow\;df=0
$$
Therefore, the kernel of the Lie algebra homomorphism
$$
\pi:\cta{M}{\alpha}\To dyn(M,\alpha)
$$
is $\set{f\cdot\eta\;|\;f=const\in\Real}\equiv\Real\cdot\eta$. So, we have the following short
exact sequence:
$$
0\;\To\;\Real\cdot\eta\;\hookrightarrow\;\cta{M}{\alpha}\;\stackrel{\pi}{\To}\;dyn(M,\alpha)\;\To\;0
$$
Assume that the $M$ is a compact manifold, and introduce a scalar product on the space of
complex-valued smooth functions on the manifold $M$ as follows:
$$
\scalar{\phi}{\psi}=\int\limits_M\phi\overline{\psi}\cdot\alpha\wedge\omega^n,\quad
\phi,\psi\in\smooth{M}_\Complex
$$
Let us denote the corresponding Hilbert space under $\hilb(M)$. A vector field $X$, on the
manifold $M$, defines an operator on the Hilbert space $\hilb(M)$, via the derivation:
$\varphi\mapsto X(\varphi)$.
\begin{lem}
If $X\in\cta{M}{\alpha}$, then the corresponding derivation operator $X:\hilb(M)\To\hilb(M)$ is
antisymmetric for the scalar product $\scalar{\cdot}{\cdot}$.
\end{lem}
\begin{prf}
Let us denote the volume form $\alpha\wedge\omega^n$ by $V$. For any function
$f\in\smooth{M}_\Complex$, we have the following:
$$
L_X(f\cdot V)=i_X\big(\underbrace{d(f\cdot V)}_0\big)+
d\big(i_X(f\cdot V)\big)=d\big(i_X(f\cdot V)\big)
$$
On the other hand, the operator of Lie derivation is a differential operator of degree 0, that is:
$L_X(f\cdot V)=X(f)\cdot V+f\cdot L_X(V)$. But by definition of the Lie algebra \cta{M}{\alpha},
we have that $L_X(V)=0$. Therefore, we obtain:
$$
X(f)\cdot V=d\big(i_X(f\cdot V)\big)
$$
If $f=\phi\overline{\psi}$, for some $\phi,\psi\in\hilb(M)$, we have that
$$
\scalar{X(\phi)}{\psi}+\scalar{\phi}{X(\psi)}=
\int\limits_Md\big(i_X(\phi\overline{\psi}\cdot V)\big)=0\;\Leftrightarrow\;
\scalar{X(\phi)}{\psi}=-\scalar{\phi}{X(\psi)}
$$
i.e., the operator $X:\hilb(M)\To\hilb(M)$ is antisymmetric.
\end{prf}

To summarize, we can state that we have a representation of the Poisson algebra of basic
functions on the contact manifold $(M,\alpha)$, in the Lie algebra of antisymmetric operators
on the Hilbert space $\hilb(M)$:
$$
\basef\ni f\mapsto\widehat{f}:\hilb(M)\To\hilb(M)
$$
It is clear that the Hilbert space $\hilb(M)$ is too large for this representation, to be
irreducible. Let us make one ``small'' step to the reduction of this representation. For any
$h\in\Real$, let
$$
\hilb_h(M)=\set{\varphi\in\hilb(M)\;|\;\eta(\varphi)=ih\cdot\varphi}
$$
As any operator of the type $\widehat{f}$ commutes with $\eta$, the subspace $\hilb_h(M)$ is
invariant for the operators $\widehat{f},\;f\in\basef$. For such operators, on the subspace
$\hilb_h(M)$, we have $\widehat{f}(\phi)=(f\cdot\eta+\ham{df})(\phi)=ihf\phi+\ham{df}(\phi)$.
The operator $\widehat{f}$ consists of the vertical and horizontal parts: $f\cdot\eta$ and
$\ham{df}$. As smaller the absolute value of the real number $h$ is, as closer the operator
$\widehat{f}$ is to its horizontal part on the space $\hilb_h(M)$.

The homomorphism from the Poisson algebra $\basef$ to the Lie algebra \cta{M}{\alpha}: $f\mapsto\widehat{f}$,
is not a prequantization, because the operators $\widehat{f}$ are antisymmetric ant not Hermitian.
To ``correct'' this situation, we can multiply them by some imaginary complex number:
$o(f)=ih\cdot\widehat{f}$. After this, redefine the commutator on the set $ih\cdot\cta{M}{\alpha}$,
so that the obtained bracket be a Lie algebra structure:
$$
\{o(f),o(g)\}=\frac{1}{ih}[o(f),o(g)]=ih[\widehat{f},\widehat{g}]
$$
After this, the pair $\big(ih\cdot\cta{M}{\alpha},\{\;,\;\}\big)$ is a Lie algebra and the mapping
$f\mapsto o(f)$ is a Lie algebra homomorphism. On the Hilbert space $\hilb_h(M)$, the operator
$o(f)$, is $o(f)=f-\frac{i}{h}\ham{df}$, where $f$ denotes the ''mutiplication by $f$`` operator:
$\varphi\mapsto f\cdot\varphi$.

In ``good'' cases, when the contact structure is regular, i.e., the space of orbits of the canonical
vector field $\eta$ is separable, the contact manifold $M$ can be considered as the total space of
a principal bundle over the space of orbits of the canonical vector field $\eta$. The structure
group of this bundle is the circle $S^1\cong U(1)$, and the form $A=2\pi i\alpha$ is a connection
form on this principal bundle. The base of this bundle, is a symplectic manifold, with symplectic
form $dA$ (see \cite{BoothbyWang}). The Hilbert space $\hilb_h(M),\;h\in\Real$ is canonically
isomorphic to the space of sections of the associated complex line bundle, via the representation
of the group $U(1)$: $\varphi_h:U(1)\To U(1),\quad\varphi_h(a)=a^h$.
\newpage
\bibliographystyle{amsplain}

\end{document}